\documentclass[a4paper,11pt]{article}

\pdfoutput=1               
\usepackage{jcappub}       
\usepackage[T1]{fontenc}   

\usepackage{graphicx}
\usepackage{slashed}

\def\sfrac#1#2{{\textstyle{#1\over #2}}}
\newcommand{\be}{\begin{equation}}
\newcommand{\ee}{\end{equation}}
\newcommand{\ba}{\begin{array}}
\newcommand{\ea}{\end{array}}
\newcommand{\bea}{\begin{eqnarray}}
\newcommand{\eea}{\end{eqnarray}}
\newcommand{\sss}{\scriptscriptstyle}
\newcommand{\R}{{\sss R}}
\newcommand{\A}{{\sss A}}
\newcommand{\W}{{\sss W}}
\newcommand{\Z}{{\sss Z}}
\newcommand{\B}{{\sss B}}
\renewcommand{\S}{{\sss S}}
\renewcommand{\L}{{\sss L}}

\def\sfrac#1#2{{\textstyle{#1\over #2}}}
\def\lsim{\mathrel{\raise.3ex\hbox{$<$\kern-.75em\lower1ex\hbox{$\sim$}}}}
\def\gsim{\mathrel{\raise.3ex\hbox{$>$\kern-.75em\lower1ex\hbox{$\sim$}}}}

\title{\boldmath Electroweak baryogenesis and dark matter from a singlet Higgs}

\author[a]{James M.\ Cline,}
\author[b,c]{Kimmo Kainulainen,}

\affiliation[a]{Department of Physics, McGill University, \\ 
                3600 Rue University, Montr\'eal, Qu\'ebec, Canada H3A 2T8}
\affiliation[b]{Department of Physics, P.O.Box 35 (YFL), \\ 
                FI-40014 University of Jyv\"askyl\"a, Finland}
\affiliation[c]{Helsinki Institute of Physics, P.O.~Box 64,\\ 
  	            FI-00014 University of Helsinki, Finland.}

\emailAdd{jcline@physics.mcgill.ca}
\emailAdd{kimmo.kainulainen@jyu.fi}

\abstract{If the Higgs boson $H$ couples to a singlet scalar $S$ via $\lambda_m
|H|^2 S^2$, a strong electroweak phase transition can be induced
through a large potential barrier that exists already at zero
temperature. In this case properties of the phase transition can be
computed analytically.  We show that electroweak baryogenesis can be
achieved using CP violation from a dimension-6 operator that couples
$S$ to the top-quark mass, suppressed by a new physics scale that can
be well above 1 TeV.  Moreover the singlet is a dark matter candidate
whose relic density is $\lesssim 3$\% of the total dark matter
density, but which nevertheless interacts strongly enough with nuclei
(through Higgs exchange) to be just below the current XENON100 limits.
The DM mass is predicted to be in the range $80-160$ GeV. }

\begin{document}
\maketitle
\flushbottom

\section{Introduction}

New physics is required for viable electroweak baryogenesis and for
dark matter (DM). A natural scenario for potentially achieving both
simultaneously is scalar dark matter $S$ that couples to the Higgs
field $H$ through the renormalizable interaction $\lambda_m |H|^2
S^2$. The case where the new scalar is another SU(2)$_L$ doublet (the
inert doublet model model \cite{Ma:2006km}-\cite{Arhrib:2012ia}) has
recently been studied in this context
\cite{Chowdhury:2011ga}-\cite{Gil:2012ya}, where working examples were
found, at the expense of some tuning of parameters in order to
sufficiently suppress the relic density to satisfy direct detection
constraints while getting a strong electroweak phase transition
(EWPT).\footnote{Specifically, the DM should be close to half of the
Higgs mass in order to have resonant annihilation $SS\to f\bar f$
through Higgs exchange in the $S$ channel, and several large quartic
couplings should combine to give a small effective coupling
$\lambda_m$.}   The case of singlet $S$ dark matter
\cite{McDonald:1993ex}-\cite{Mambrini:2011ik} has also been considered
with respect to its impact on the electroweak phase transition in
refs.\ \cite{{McDonald:1993ey}}-\cite{Gonderinger:2012rd} (for the
effect of singlets on the EWPT without requiring them to be dark
matter, see \cite{Anderson:1991zb}-\cite{Espinosa:2011eu}). It was
found that with a real singlet, one could either enhance the EWPT or
have dark matter, but not both simultaneously.  One of our main points
is that a real singlet is adequate if one relaxes the requirement that
$S$ be the dominant dark component.  We will show that $S$ could
constitute up to $3$\% of the total DM mass density, yet still be
relevant for direct detection because of its large scattering cross
section on nuclei. (For other references about the possible EWPT/DM
connection, see \cite{Dimopoulos:1990ai}-\cite{Carena:2011jy}.)

Because of the complexity of the finite-temperature effective potential that controls the
dynamics of the EWPT, it is often desirable to do a fully numerical analysis, as for example in ref.\ \cite{Cline:2011mm}.  But recently it was emphasized
\cite{Espinosa:2011ax,Espinosa:2011eu} that in some situations where the phase transition is strong, analytic methods are applicable, which greatly simplify the search for working
models.  This is the case when there is a large barrier already at tree level between the
electroweak symmetry  breaking vacuum $\langle H\rangle = v_0/\sqrt{2}, \langle S\rangle = 0$, and a nearly degenerate symmetric one with $\langle H\rangle = 0, \langle S\rangle = w_0$.  Then the transition can be triggered by rather weak thermal corrections, at a temperature significantly lower than the critical vacuum expectation value of the Higgs,  $v_c$, which can be relatively close to the zero-temperature one $v_0$. We take advantage of this in the present study, although the solution of transport equations needed to compute the baryon asymmetry must still be done numerically.

In addition to promoting $S$ to a dark matter candidate, we take advantage of it to get the CP-violation required for baryogenesis by introducing a dimension-6 
operator, that modifies the top quark mass at nonzero $S$.  The full mass term takes the form
\begin{equation}
	y_t \bar Q_\L H \left(1+ {\eta\over\Lambda^2}S^2\right)t_\R +{\rm h.c.}
\label{dim6op}
\end{equation}
where $\eta$ is a complex phase and $\Lambda$ is a new 
physics scale. During the EWPT, the top quark mass thus gets a 
spatially-varying complex phase along the bubble wall profile, 
which provides the source of CP violation needed to generate the 
baryon asymmetry.  Ref.\ \cite{Espinosa:2011eu} considered the 
analogous dimension-5 operator involving $S/\Lambda$, but here we 
are forced to use $S^2/\Lambda^2$ because of the $Z_2$ symmetry
$S\to -S$ needed to prevent decay of $S$, as befits a dark matter candidate.

We review the method of construction of the effective potential in
section \ref{effpot}, constraints from invisible Higgs decays
in section \ref{higgsbound}, and direct detection constraints on the scalar
dark matter candidate in section \ref{ddconst} along with some results
from a random scan over model parameters. The absence of other
constraints on the model is explained in section \ref{other}.
The computation and
resulting distributions of value for the baryon asymmetry are
described in section \ref{bau}. Conclusions are given in section
\ref{conc}.

%
\section{Effective potential}
\label{effpot}
%

We follow refs.\ \cite{Espinosa:2011ax,Espinosa:2011eu}, starting from the tree-level potential for the Higgs doublet $H$ and real singlet $S$,
\begin{eqnarray}
	V_0 &=&  \lambda_h\left(|H|^2 -\sfrac12 v_0^2\right)^2 +
	 \sfrac14\lambda_s\left( S^2 - w_0^2\right)^2 
	 \nonumber\\
	 &+& \sfrac12\lambda_m |H|^2 S^2 \,.
\end{eqnarray}
This potential has the $Z_2$ symmetry $S\to -S$ that is needed to guarantee the stability of $S$ as a DM particle, but parameters can be chosen such that the $Z_2$ breaks spontaneously at high temperatures, giving $S$ a VEV (with $H=0$) in the electroweak symmetric vacuum, while the true vacuum is along the $H$ axis at $T=0$.\footnote{Ref.~\cite{Espinosa:2011eu} notes that domain walls associated with this spontaneous breaking of $Z_2$ would only come to dominate the energy density of the universe at low temperatures $T\sim 10^{-7}$ GeV; but by this time the symmetry is restored and the domain walls are no longer present.}  The finite-temperature effective potential for the real fields $H=h/\sqrt{2}$ and $S$ can be written in the form
\begin{eqnarray}
	V &=& {\lambda_h\over 4}\left(h^2-v_c^2 +{v_c^2\over w_c^2}S^2\right)^2
	+ {\kappa\over 4}S^2 h^2\nonumber\\
	&+& \sfrac12(T^2 - T_c^2)(c_h h^2 + c_s S^2) \,,
\label{Veff}
\end{eqnarray}
where the parameter $w_0$ has been traded for its counterpart $w_c$ at the critical temperature of the phase transition $T_c$, $v_c$ is the corresponding critical VEV of $h$, and the following relations hold:
\begin{eqnarray}
	\kappa &\equiv& \lambda_m -2\lambda_h{v_c^2\over w_c^2}\\
 	T_c^2 &=& {\lambda_h\over c_h}\left(v_0^2-v_c^2\right) \,.
\end{eqnarray}
Here the coefficients $c_h$ and $c_s$ determine the $O(T^2)$ corrections to the  masses of $h$ and $S$, and are given in terms of the gauge and other couplings by
\begin{eqnarray}	
	c_h &=& \sfrac{1}{48}\left(9g^2 + 3g'^2 + 12y_t^2 
	 + \lambda_h\left(24 + 4\frac{v_c^2}{w_c^2} \right) + 2\kappa\right)
\nonumber\\
	c_s &=& \sfrac{1}{12}\left( \lambda_h\left(3{v_c^4\over w_c^4}
	                  + 4\frac{v_c^2}{w_c^2} \right) + 2\kappa\right) \,,
\end{eqnarray}
while the zero-temperature masses are given by
\begin{eqnarray}
 	 m_h^2 &=& 2\lambda_h v_0^2,\\
	 m_\S^2 &=& \sfrac12\kappa v_0^2 + \lambda_h(v_0^2-v_c^2)\left({v_c^2\over w_c^2}-
	{c_s\over c_h}\right) \,.
\end{eqnarray}

Counting parameters, and taking the Higgs mass to be determined as $m_h = 125$ GeV 
\cite{ATLAS:2012ae}-\cite{{CMS:2012gu}}, one sees that a given model can be specified by choosing three of them freely, which we take to be $\lambda_m$, $v_c/w_c$ and $v_0/v_c$.  It can be shown that as long as $v_0/v_c>1$ so that $T_c$ exists, the temperature at which the $Z_2$ spontaneously breaks (so that $S$ gets a VEV prior to the EWPT) is always higher than $T_c$. There is also a restriction 
\begin{equation}
\frac{v_c^2}{w_c^2} > \frac{c_s}{c_h} 
\label{eq:consistency}
\end{equation}
ensuring that the electroweak breaking minimum is the lower of the two at zero temperature.  More general discussions of the impact of dark matter interacting with the Higgs on stability of the electroweak vacuum and perturbativity of the couplings have been given in  references \cite{Gonderinger:2012rd,Cheung:2012nb}.  In the present
work we will enforce perturbativity of $\lambda_m$ at the weak scale by restricting the range over which it varies to be less than unity in our Monte Carlo scans.  We will not make any assumptions about the scale at which the effective theory must be UV-completed in order to deal with possible Landau poles, reflecting breakdown of perturbativity
at higher scales. 

One can efficiently scan the parameter space for models with a strong EWPT characterized by\footnote{for caveats concerning the reliability of this estimate, see references 
\cite{Patel:2011th,Morrissey:2012db}}
\begin{equation}
  \frac{v_c}{T_c} > 1 \,,
\label{eq:sphaleronbound}
\end{equation}
which is needed to satisfy the sphaleron washout constraint, by randomly varying the inputs  $\lambda_m$, $v_c/w_c$ and $v_0/v_c$ over reasonable ranges. This procedure easily generates many examples of models with a strong phase transition. However we would like to combine this with the requirement that $S$ provides a viable dark matter candidate, which we discuss in the next section.

In addition to the potential (\ref{Veff}) considered in \cite{Espinosa:2011ax,Espinosa:2011eu}, there is a correction coming from the dimension-6 operator (\ref{dim6op}) due to its contribution to the finite-$T$ top quark mass,
\begin{equation}
	\delta V = {y_t^2\over 8}T^2 h^2 (S/\Lambda)^4
\label{dVeff}
\end{equation}
at leading order in the high-$T$ expansion.  It does not change the critical temperature as defined in (\ref{Veff}) because it is higher than quadratic order in the fields, and it also does not change the positions of the critical VEVs since it vanishes, along with its first derivatives, on either field axis.  The main effect of this operator then is to increase the height of the barrier at the critical temperature (making the bubble wall thinner), and probably to bend the path of the bubble wall in field space (to be discussed in section \ref{bau}).  These aspects of the problem are being treated in a rather rough way already, and we further require that $S/\Lambda$ remain small for consistency, so we do not expect (\ref{dVeff}) to play an important role, and thus omit it from our analysis.

The above procedure implicitly assumes that the high-temperature expansion is valid for the models of interest.  This is an assumption that can only be checked by redoing the analysis using the full effective potential, which could be an interesting undertaking for future work, but is beyond the scope of the present paper.  We note however that precisely such a check was done in ref.~\cite{Espinosa:2011ax} to justify this approximation for a general class of singlet Higgs extensions to the standard model.  There it was found that although the high-$T$ expansion tends to somewhat overestimate the strength of the phase transition, nevertheless cases with $v_c/T_c>1$ were always present under the more exact
treatment, for models nearby in parameter space to those favored by the approximate formalism.

%
\section{Invisible decay of Higgs boson}
\label{higgsbound}
%

If $m_\S<m_h/2$, the $S$-particle would contribute to the invisible decay width of the Higgs boson. 
The current constraint on the branching to invisible channels 
is ${\rm Br}_{H \to \rm invis} \lsim 0.35$ \cite{Espinosa:2012im}. 
(The exact value ranges between 0.32 and 0.37 for $m_h = 125-126.5$;
our results are not sensitive to this variation.)
Since the Higgs decay width to visible channels at $m_h=125$ GeV 
is $\Gamma_h = 4.07$ MeV this implies
\begin{equation}
\Gamma_{H \to SS} < 2.2 \,\rm MeV \,,
\label{BRconstraint}
\end{equation}
while the predicted the decay width for $H\to SS$ is
\begin{equation}
\Gamma_{H \to SS} = \frac{\lambda_m^2v_0^2}{32 \pi m_h} (1-4m_\S^2/m_h^2)^{1/2} \,.
\label{GammaHSS}
\end{equation}
Using (\ref{GammaHSS}) we can recast (\ref{BRconstraint}) as
\begin{equation}
\lambda_m \lsim 0.051 \;  \left(\frac{\rm GeV}{\sfrac12 m_h-m_\S}\right)^{1/4} \,.
\label{BRconstraint2}
\end{equation}
This constraint is strong enough to effectively  exclude all interesting models with $m_\S < m_h/2$.  In the following, we will show that $\lambda_m$ can be no smaller than $\sim 0.1$ to get a strong phase transition. For such values (\ref{BRconstraint2}) 
then implies $\sfrac12 m_h - m_\S \lesssim 70$ MeV. It will be seen below (see for example fig.\ \ref{fpan}) that there is a narrow, finely-tuned region of parameter space where $m_\S \cong \sfrac12 m_h$, corresponding to resonant annihilations of the dark matter in order to achieve low enough relic density.  This restriction therefore only serves to make this region a bit narrower than it would otherwise have been.

\begin{figure}[t]
\hspace{-0.4cm}
\centerline{
\includegraphics[width=0.65\hsize]{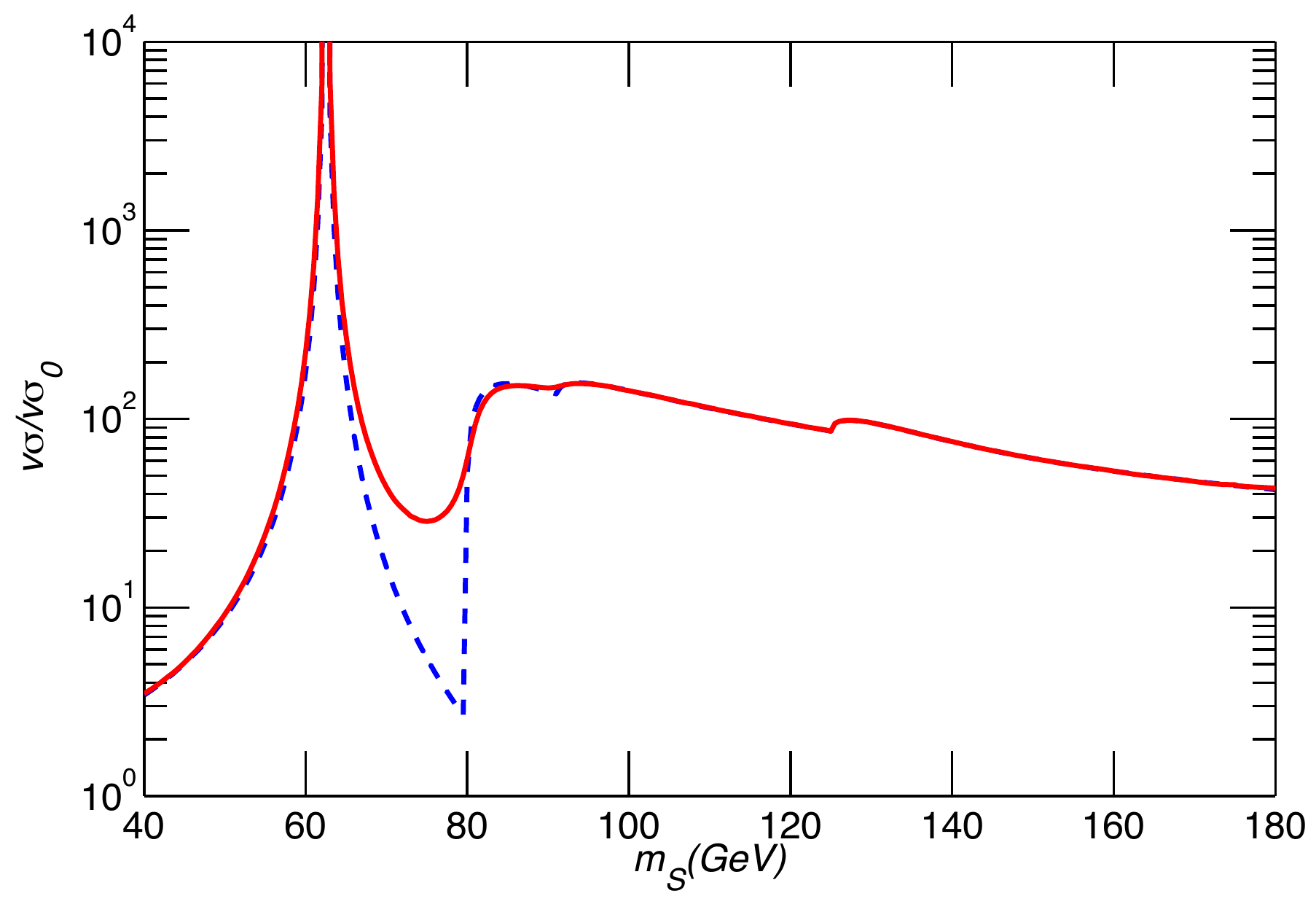}}
\caption{Comparison of the $SS$ annihilation cross section 
(in units of the standard relic density value 
$\langle\sigma v\rangle_0$) as a function of $m_\S$ using the 
approximations (\ref{twobody}) (dashed, blue) and (\ref{better}) 
(solid, red), for the case $\lambda_m=0.5$. }
\label{sigcomp}
\end{figure}

\bigskip
%
\section{Dark matter constraints}
\label{ddconst}
%

Because of the $h^2 S^2$ cross-coupling, $S$ acquires the interaction $\lambda_m v_0 h S^2$ after electroweak symmetry breaking, and so scatters from nuclei by Higgs exchange.
Its relic abundance is partially determined by Higgs-mediated
 annihilations $SS\to b\bar b$, but if $m_\S > m_{\Z,\W}$ or $m_h$, 
there are more important contributions from $SS \to WW,\,ZZ,\,hh$.  Defining $r_i = m_i^2/m_\S^2$, the respective contributions to the 
annihilation cross section from these processes at $s=4m_\S^2$ 
are given by
\begin{eqnarray}
\label{twobody}
	\langle\sigma v\rangle_i &=& {\lambda_m^2\over 8\pi m_\S^2\left[(4-r_h)^2 +
	r_h\Gamma_h^2/m_\S^2\right]} \\
	&\times&\left\{ \begin{array}{ll} 6 r_b (1-r_b)^{3/2},& b\bar b\\
	& \\
	\delta_\A {r_\A^2}\left(2 + (1-2/r_\A)^2\right)\sqrt{1-r_\A},& AA\\
	& \\
	2\left({\lambda_m\over\lambda_h}\, 
	{(1-r_h/4)r_h\over r_h-2} + 1+{r_h\over 2}\right)^2\sqrt{1-r_h},&
	hh
	\end{array}\right.\nonumber
\end{eqnarray}
where $A$ stands for $W$ or $Z$, and 
$\delta_\W = 1$, $\delta_\Z = 1/2$. However, near $m_\S \approx m_{\Z,\W}$ these do not include the contributions from 4-body final states due to virtual $W$ and $Z$ emission.  A more accurate expression for the contributions from all final states except for $hh$ in this region is given by factorizing into  the $SSh$ fusion part times the virtual $h$ decay using the full width of the Higgs,
\begin{equation}
	\langle\sigma v\rangle_{\slashed{h}} = 
	\left.{2\lambda_m^2 \Gamma_h(s)v_0^2\over \sqrt{s} 
	      \left[(s-m_h^2)^2 + m_h^2\Gamma_h^2\right]}
	\right|_{s=4m_\S^2}
\label{better}
\end{equation}
The $hh$ final state contribution is added to this as in 
(\ref{twobody}). In the range 80 GeV$< \sqrt{s}<$340 GeV we take 
$\Gamma_h(\sqrt{s})$ from ref.\ \cite{Dittmaier:2011ti}, which also
includes QCD corrections. Above $340$ GeV we revert to (\ref{twobody}),
because here the 1-loop Higgs self-interaction corrections, also
included in \cite{Dittmaier:2011ti} as for $m_h=\sqrt{s}$, would be
too large for a light Higgs with $m_h=$ 125 GeV. The resulting
annihilation cross sections in these approximations are compared in
Fig.\ \ref{sigcomp}, showing that the inclusion of the 4-body final
states gives a more smooth dependence on $m_\S$ near the $WW$
threshold.

\begin{figure}[t]
\hspace{-0.4cm}
\centerline{
\includegraphics[width=9.5cm]{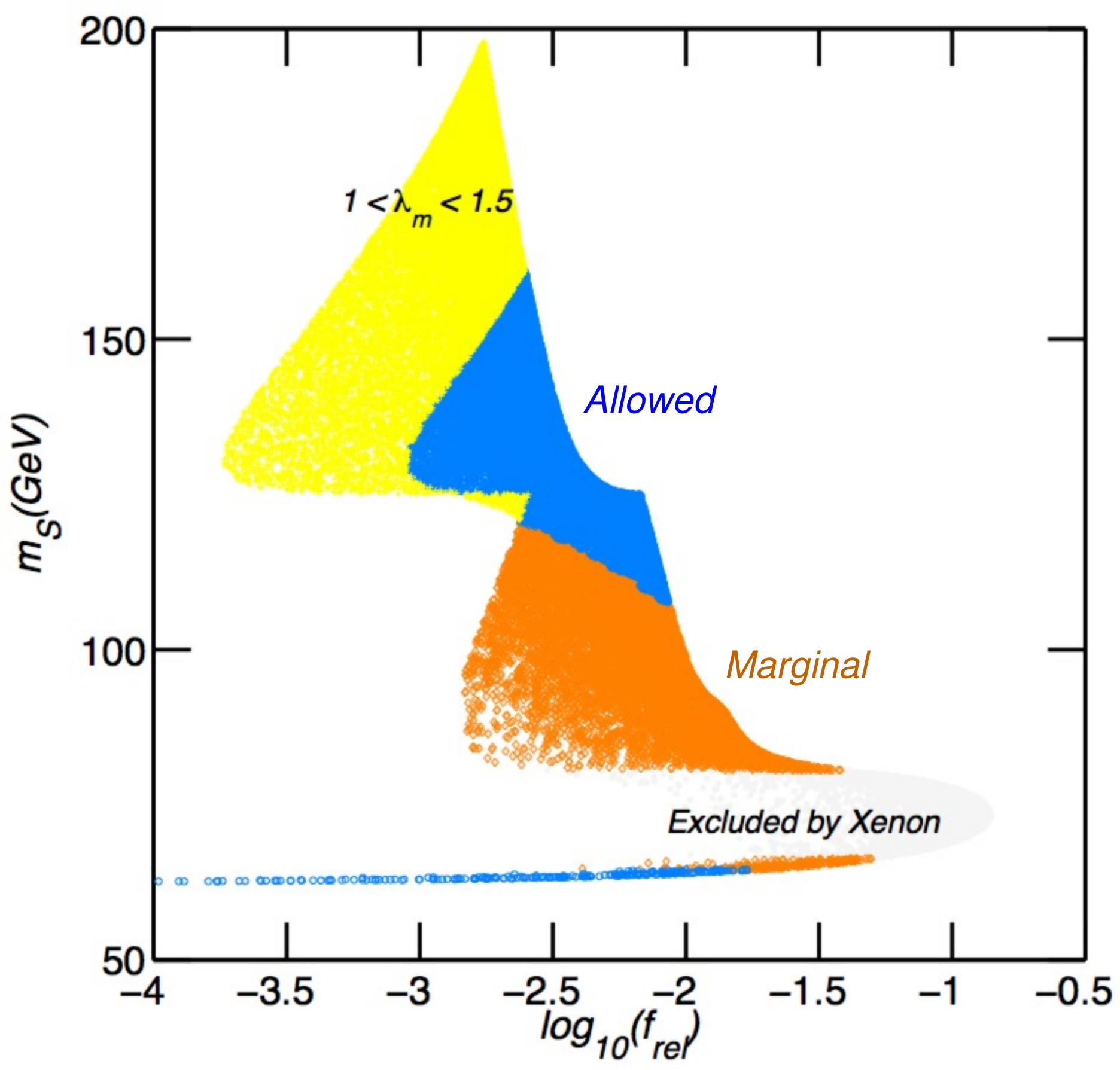}}
\caption{Scatter plot $f_{\rm rel}$ vs.\ $m_{S}$ from a random scan of
parameter space with input parameters varying in the ranges
$\lambda_m=0.1-1$, $v_0/v_c = 1.1-10$, $\log_{10} v_c/w_c =
(-1)-(+1)$. Different groups of points are distinguished by their
relation to XENON bound: lightest gray points (filled circles) are
already excluded, orange diamonds (``marginal'') are uncertain and blue
circles (``allowed'') are still allowed even by strongest XENON limit.
Yellow plus signs  show the extension of the allowed
region when the upper bound on $\lambda_m$ is pushed to 1.5.}
\label{fpan}
\end{figure}

Often an approximation is made
by requiring that the total cross section 
$\langle\sigma v\rangle = \sum_i \langle\sigma v\rangle_i$ be
 equal to the standard value 
$\langle\sigma v\rangle_0 = 3\times 10^{-26}$cm$^3$/s to 
get the correct relic density from thermal freezeout. We find however, 
in agreement with~\cite{Steigman:2012nb}, that this procedure may 
overestimate relic density by as much as 30-40\% in 
the range of interest. We thus adopt the more accurate but still quite 
efficient freeze-out formalism described for example in refs.~\cite{Enqvist:1990yz,Enqvist:1988dt}. 

To quantify the DM abundance we define the ratio
\begin{equation}
	f_{\rm rel} = {\Omega_\S h^2 \over 0.11} \,.
\end{equation}
In the Monte Carlo scan over models that 
that selects for a strong EWPT (to be discussed in greater detail
below), 
we find that none of them can give $f_{\rm rel}\cong 1$ while
remaining consistent with the direct detection constraints
(see figure \ref{fpan}). However, there is no need for $S$ to be the 
only DM component; it could make a subdominant contribution to the 
total DM density, while still interacting strongly enough with 
nuclei to be potentially detectable.  If $f_{\rm rel} < 1$, then the 
relic density of $S$ is suppressed relative to the observed value by 
$f_{\rm rel}$, and larger values of the coupling $\lambda_m$, 
which controls the barrier height and thus the strength of the phase 
transition, become allowed. We will exploit this possibility
in the following.

\begin{figure}[t]
\hspace{-0.4cm}
\centerline{
\includegraphics[width=0.65 \hsize]{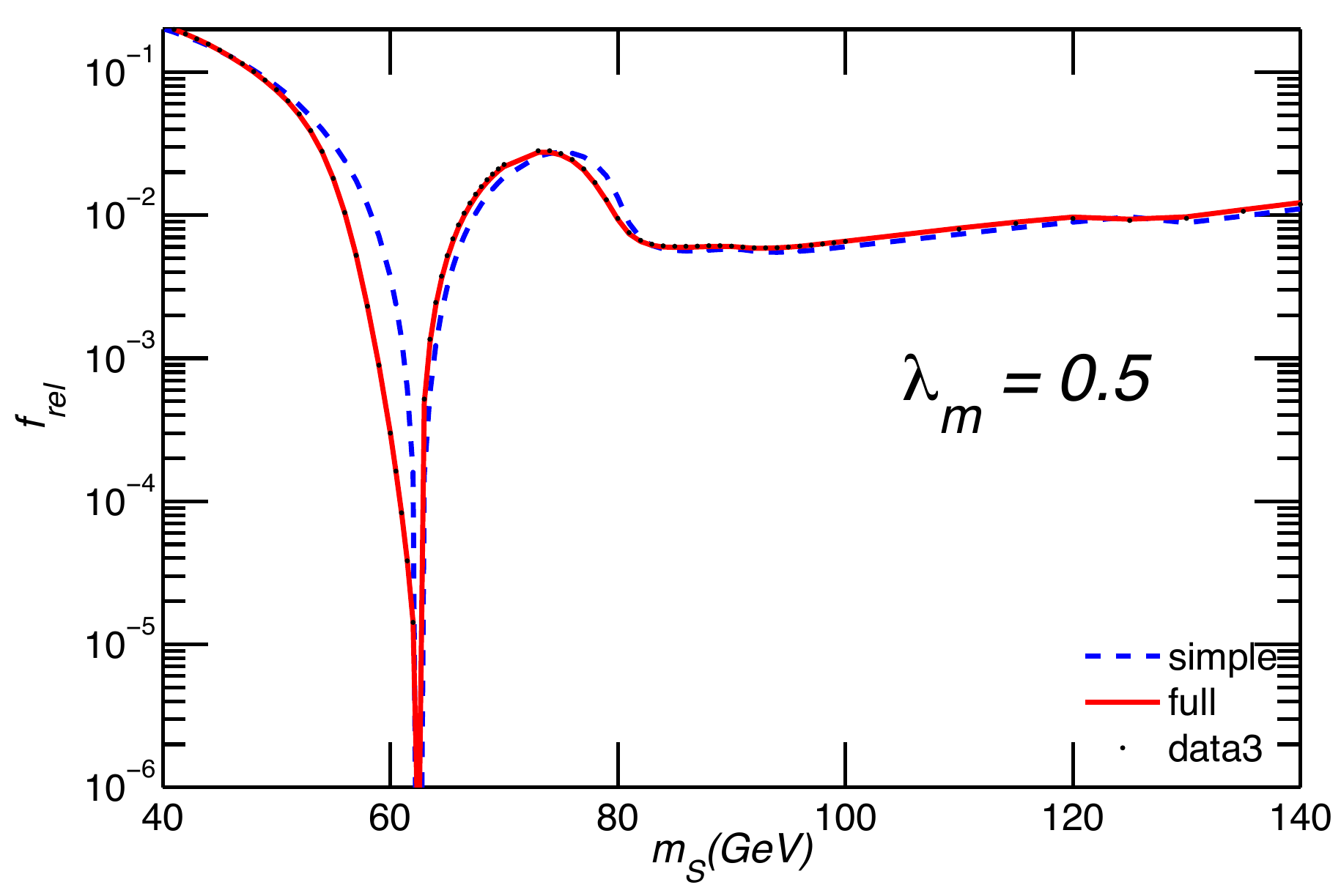}}
\caption{Plot of $f_{\rm rel}$ as a function of $m_\S$ in the case $\lambda_m=0.5$. Blue dashed line shows the calculation using $\langle \sigma v \rangle_{s=4m_\S^2}$ and red solid line the one using the accurate thermally averaged annhilation cross section.}
\label{relicdensity}
\end{figure}

Before coming to constraints from direct detection, we point out a
subtlety in the DM abundance determination. Because the Higgs
resonance is very narrow, the cross section at the pole is quite
large, and the annihlation cross section evaluated at $s=4m_\S^2$ is
not an accurate approximation to the thermally averaged cross section
when $m_\S \lsim m_h/2$, where the WIMP thermal
distribution may be overlapping with the pole.  The effect of using
the accurate thermal annihilation cross section (see
ref.~\cite{Gondolo:1990dk}) is illustrated in
figure \ref{relicdensity}. In fact neither of the results shown 
there can
be trusted in the affected range 50 GeV $\lsim m_\S < m_h/2$. 
Whereas $\langle \sigma v\rangle_{s=4m_\S^2}$ is an underestimate, the
integrated $\langle \sigma v\rangle$ is an overestimate for the
annihilation rate, because in reality only a narrow range of momenta
see the pole at any given time and rapid kinetic-equilibrium-restoring
processes would be needed to continuously feed new states into this
range. We bracketed the uncertainty by performing the computation both
ways. In the end this does not affect our results because
all models sensitive to the effect turn out to be excluded by the invisible Higgs
decay constraint (\ref{BRconstraint2}), regardless of the way in which
the
relic density was computed.

Next we consider the constraints from direct detection.  The cross section for spin-independent scattering on nucleons by virtual Higgs exchange is given by \cite{Barbieri:2006dq}
\begin{equation}
 \sigma_{\sss SI} = \frac{\lambda^2_{m}f_N^2}{4\pi}\frac{\mu^2 m^2_n}{m^4_h m^2_s}
\label{sigma_dd}
\end{equation}
where $\mu = m_n m_{\S}/(m_n+m_\S)$ is the DM-nucleon reduced mass.  The Higgs-nucleon
coupling $f_N$ suffers from hadronic uncertainties, which have been estimated to be as large as $f_N=0.26-0.63$ \cite{Mambrini:2011ik}.  However recent lattice studies obtain roughly consistent values $f_N\cong 0.33-0.36$ \cite{Toussaint:2009pz,Young:2009zb}.  For definiteness we adopt the value $f_N=0.35$ which also agrees with ref.\ \cite{Ellis:2000ds}.

\begin{figure}[t]
\hspace{-0.4cm}
\centerline{
\includegraphics[width=0.9 \hsize]{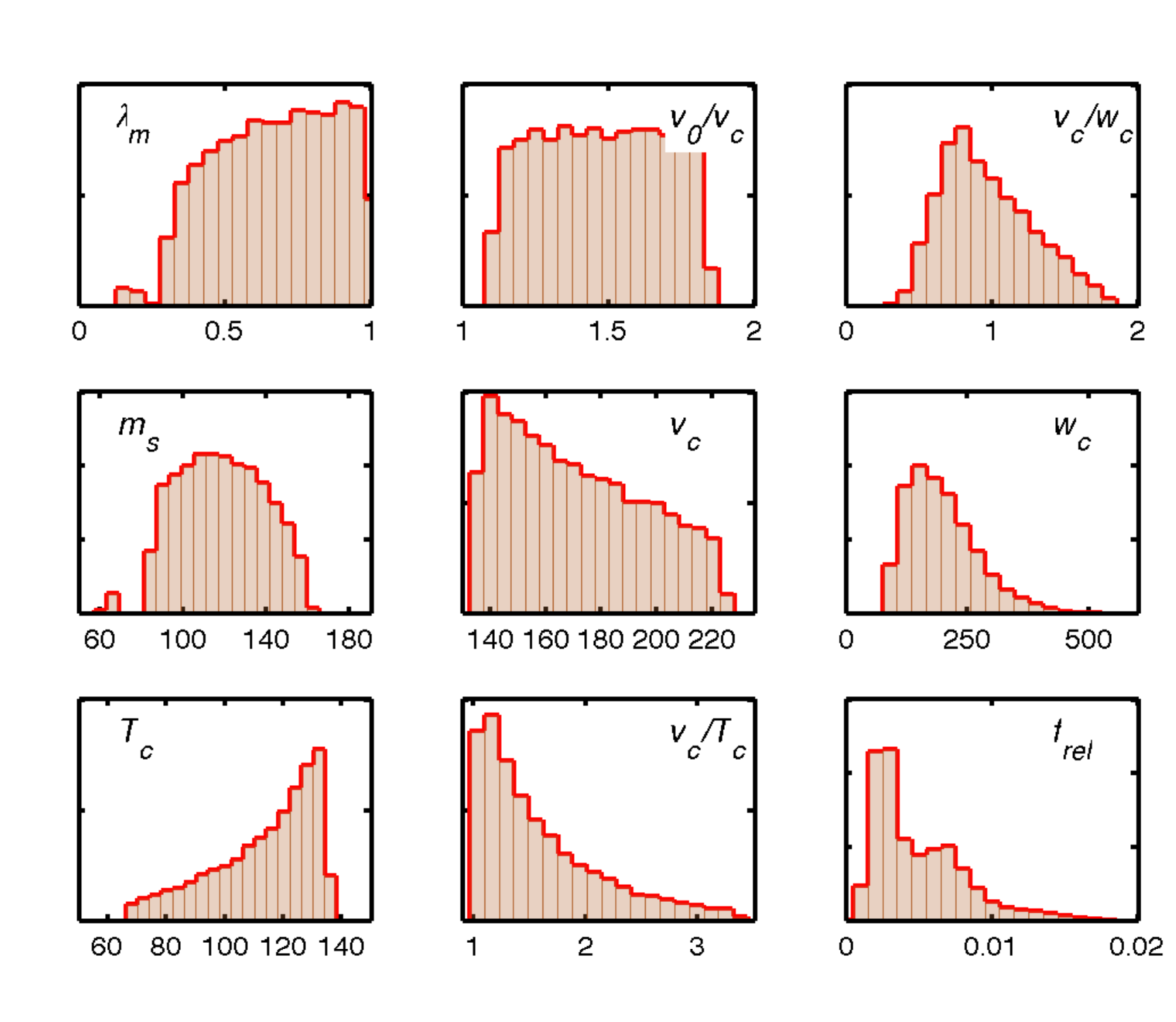}}
\caption{Distributions of parameters satisfying the constraints (\ref{eq:consistency}, \ref{eq:sphaleronbound}, \ref{BRconstraint}) and the nominal DM direct detection bound~(\ref{Xeconst}).  
Top row shows input parameters, bottom two rows are derived.  Dimensionful quantities are in GeV units.}
\label{panel-sm}
\end{figure}

The XENON100 experiment currently sets the strongest bounds on
$\sigma_{\sss SI}$ \cite{Aprile:2012nq} in the range of $m_\S$ that is of interest for the EWPT.  We would like to use their limit to constrain the ratio $f_{\rm rel}$.  That limit of course is based upon the assumption $f_{\rm rel}=1$, and becomes proportionally weaker if $f_{\rm rel}<1$.  We define the fraction
\begin{equation}	
	f_{\rm Xe} = {\sigma_{\rm Xe}\over \sigma_{\sss SI}}
\end{equation}
where $\sigma_{\rm Xe}$ is the XENON100 90\% c.l.\ upper limit for
standard DM of a given mass, and $\sigma_{\sss SI}$ is the predicted cross section (\ref{sigma_dd}). Then the constraint on subdominant DM becomes
\begin{equation}
	f_{\rm rel} <  f_{\rm Xe} \,.
\label{Xeconst}
\end{equation}
We find that these bounds can be satisfied for many models that give rise to a strong EWPT.  
For example, a random sample of $2\times10^6$ models where the input 
parameters are varied over the ranges $\lambda_m=0.1-1$, 
$v_0/v_c = 1.1-10$, $\log_{10} v_c/w_c \in (-1,1)$ produces $22500$ models consistent with the constraint (\ref{Xeconst}) as well as with  the sphaleron washout bound (\ref{eq:sphaleronbound}), the consistency requirement (\ref{eq:consistency}) and the invisible Higgs decay width (\ref{BRconstraint}) of previous sections. Distributions of various parameters in this set of models can be seen in Fig.\ \ref{panel-sm}. One observes that the DM mass is typically in the range $80-160$ GeV, 
for our choice $\lambda_m < 1$. (Fig.\ \ref{fpan} illustrates that higher
masses are correlated with larger values of $\lambda_m$).
The $v_c$ values fall in the range $140-220$ GeV and as $T_c$ tends to
be around 100 GeV strong phase transitions are found with $v_c/T_c$ as
high as $3.5$. The $w_c$ distribution peaks at $w_c
\approx 160$ GeV with $w_c < 500$ GeV and the relic density fraction
$f_{\rm rel}$ tends to be $\lesssim 0.01$. 

\begin{figure}[t]
\hspace{-0.4cm}
\centerline{
\includegraphics[width=10cm]{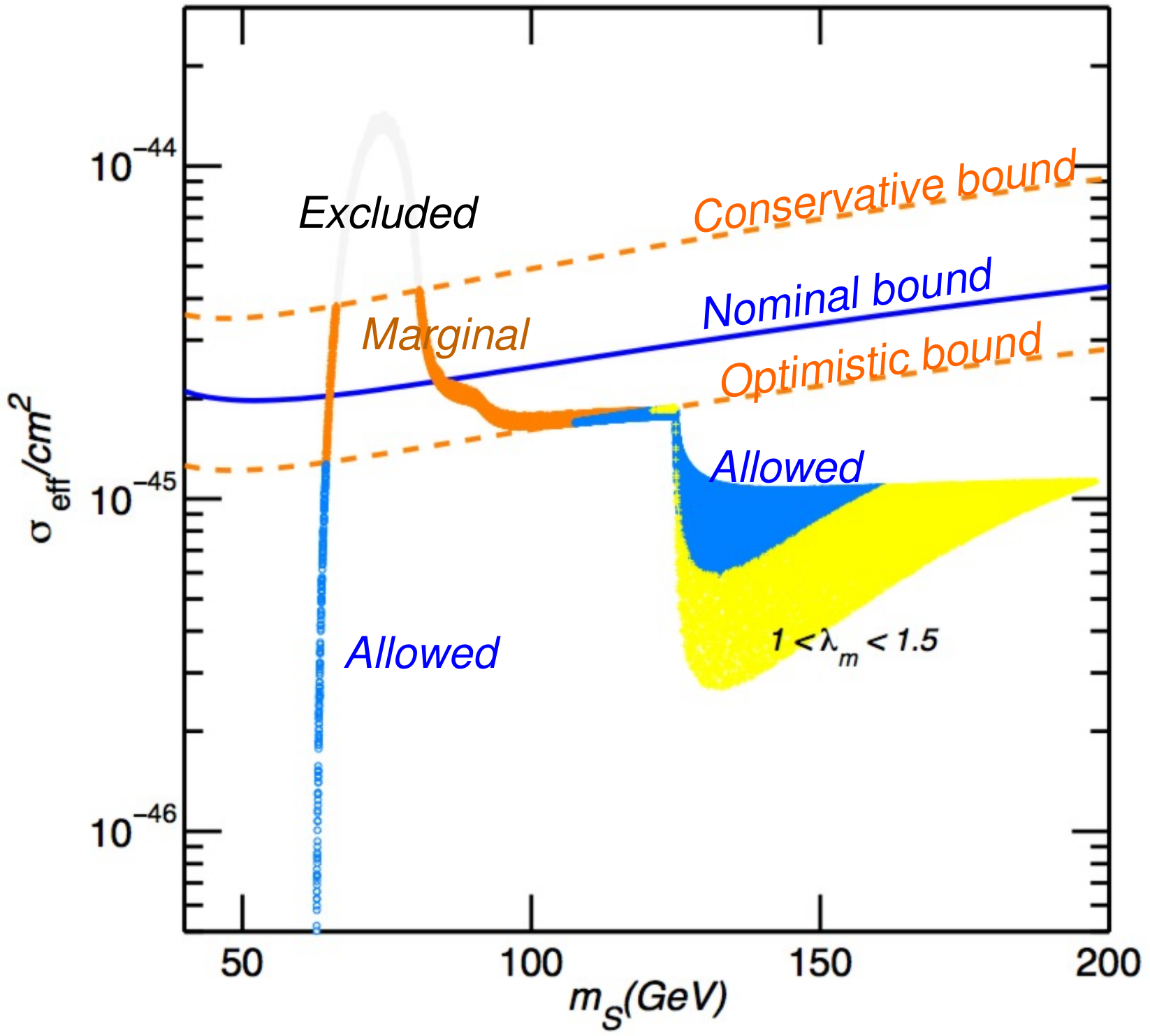}}
\caption{Scatter plot of the expected cross section 
$\sigma_{\rm eff} \equiv f_{\rm
rel}\, \sigma_{\sss SI}$ in Xenon
experiment vs.\ $m_{S}$ corresponding to the data shown in figure
\ref{fpan}, also with the same color coding. Solid blue (``Nominal
bound'')
line shows the nominal XENON100 bound \cite{Aprile:2012nq} and the dashed orange
 lines the more conservative (upper curve) and more optimistic
(lower curve) bounds reflecting uncertainties due to local DM
distribution.}
\label{fpan2}
\end{figure}

We show the scatter plot of accepted models in $f_{\rm rel}$ versus
$m_{S}$ in figure~\ref{fpan} and the same data in
figure~\ref{fpan2} as $m_\S$ versus $\sigma_{\rm eff} \equiv f_{\rm
rel}\, \sigma_{\sss SI}$.  The cross section $\sigma_{\rm eff}$
indicates
the reach of the future XENON experiments to rule out a given model,
or to verify the existence of its associated DM particle. All direct DM
bounds inevitably suffer from uncertainties in the local Galactic
abundance and velocity distribution of the DM. We estimate the effect
of these uncertainties on the latest XENON100 constraint following
ref.~\cite{Fairbairn:2012zs}, which shows that the constraint derived
from standard assumptions about the local DM distribution could
reasonably be stronger or weaker by the respective amounts
in $\Delta\log_{10}\sigma_{\rm Xe}$:
\be
\begin{array}{ll}-0.29128 + 0.4557\,x - 0.081349\, x^2,&{\rm upper}\\
-1.3329 + 1.4365\, x - 0.59337\, x^2 + 0.080595\, x^3,&{\rm lower}
\nonumber
\end{array}
\ee
where $x=\log_{10}(m_\S$/GeV).  These are our digitizations of the
90\% c.l.\ curves of fig.\ 5 [left, bottom] of
\cite{Fairbairn:2012zs}.
 The altered bounds are shown by the
dashed orange lines around the solid blue (``Nominal bound'')
line corresponding to the current XENON100 bound \cite{Aprile:2012nq},
also shown in fig.\ 
\ref{fpan2}. The colour-coding of the points in Figures
\ref{fpan},\ref{fpan2} are the same. The  gray
regions are already excluded by even the weakest of the bounds, the
orange regions (``Marginal'')  fall into the region of uncertainty and
the blue ones (``Allowed'') are still allowed. 
All of these correspond to models with $\lambda_m <1$, whereas
yellow indicates the extension of the allowed region when the upper
bound on the coupling is relaxed to $\lambda_m <1.5$. Note that the
current XENON100 results have only begun to encroach on the allowed
parameter values.  On the other hand it is interesting that most
of the models (with the exception of the rare ones where $m_\S\cong
m_h/2$) will become discoverable in the relatively near future
as direct detection sensitivity
continues to improve.

%
\bigskip
\section{Other constraints}
\label{other}
%

Because of the inert nature of the $S$ and the fact that it gets a VEV
only at high temperature, we are free from the additional constraints
that had to be imposed on similar models where the singlet Higgs was
not required to be a DM candidate.
For example, nothing prevents us from choosing the
phase $\eta$ in (\ref{dim6op}) to be maximally CP-violating. Ref.\
\cite{Espinosa:2011eu} considers the two-loop Barr-Zee contributions
to the electric dipole moments of the electron and neutron.  But this
requires $h$-$s$ mixing, which does not occur in our model. Ours is
similar to models in which CP is broken spontaneously at high
temperature in this respect.  

Because of the singlet nature of $S$ and its sole couplings being to
the Higgs (without mixing), and through the dimension-6 operator (\ref{dim6op}), there are no other direct constraints on its mass from collider searches, nor from precision electroweak observables.

%
\bigskip
\section{Baryon asymmetry}
\label{bau}
%

The baryon asymmetry depends upon a source of CP violation that biases
sphaleron interactions near the expanding bubble walls toward
production of baryons, as opposed to antibaryons.  We take our
relevant CP-violating parameter to be the phase $\eta$ in the
dimension-6 coupling in (\ref{dim6op}), and for definiteness  we fix
$\eta=e^{i\pi/2}$ to maximize the CP violation.  (Since the baryon
asymmetry $\eta_B$ goes linearly in the imaginary part, the
generalization to arbitrary phases is straightforward.)  Then inside the
bubble walls during the phase transition, the top quark has a
spatially varying complex mass, given by
\begin{equation}
	m_t(z) = {y_t\over\sqrt{2}} h(z) \left(1 + i {S^2(z)\over\Lambda^2}\right)\equiv
		|m_t(z)| e^{i\theta(z)}
\end{equation}
where $z$ is taken to be the coordinate transverse to the wall,  in
the limit that it has grown large enough to be approximated as
planar.  The existence of the nontrivial phase $\theta(z)\cong
S(z)/\Lambda$ is sufficient to source the baryon asymmetry.  In the
following, we will initially fix $\Lambda=1$ TeV for the computation
$\eta_\B$. Since $\eta_\B\sim
1/\Lambda^2$ for large $\Lambda$, one can always rescale $\Lambda$ to adjust $\eta_\B$ to
the desired value.

We follow ref.\ \cite{{Espinosa:2011eu}} in approximating the bubble wall profiles in the form
\begin{eqnarray}
	h(z) &=& \sfrac12 v_c(1+\tanh(z/L_w))\nonumber\\
	S(z) &=& \sfrac12 w_c(1-\tanh(z/L_w))
\end{eqnarray}
where the wall thickness is taken to be 
\begin{equation}
	L_w = \left({2.7\over\kappa}
	  \left( \frac{1}{w_c^{2}} + \frac{1}{v_c^{2}} \right)\left(1 +
	{\kappa w_c^2\over 4 \lambda_h v_c^2}\right)\right)^{1/2} \,.
\end{equation}
This fully determines the top quark mass profile for a given model.

The baryon asymmetry is determined by first solving transport equations for the chemical potentials and velocity perturbations of various fields that develop an asymmetry in the vicinity of the bubble wall.  We improve upon the treatment given in \cite{{Espinosa:2011eu}} by using the more recent and complete transport equations of \cite{{Fromme:2006wx}}, which are based on the semiclassical baryogenesis mechanism of refs.\cite{Joyce:1994zt,Cline:1997vk,Cline:2000nw,Kainulainen:2001cn,Kainulainen:2002th} that determine the chemical potentials of  $t_\L$, $t_\R$, $b_\L$ (the left-handed bottom quark) and $h$, rather than those of \cite{Bodeker:2004ws}.  We also correct an apparent error in \cite{{Espinosa:2011eu}} where there was a mismatch between the orientation of the bubble wall and the transport equations that were solved. (The transport equations are not symmetric under $z\to -z$ because it matters whether the wall is expanding into the symmetric phase (correct) or into the broken phase (incorrect).)

One then calculates the left-handed  baryon chemical potential
\begin{equation}
	\mu_{B_L} = \sfrac12(1+4K_{1,t})\mu_{t_L} +\sfrac12(1+4K_{1,b})\mu_{b_L}
	+2K_{1,t} \mu_{t_R}\,, 
\end{equation}
which is a linear combination of the quark chemical potentials; see Fig.\ 8 of ref.\ \cite{Cline:2011mm} for 
for a graph of the functions $K_{1,i}(m_i(z)/T)$.  The baryon asymmetry depends upon the integral of $\mu_{B_L}$  over the symmetric phase in front of the wall 
\begin{equation}
	\eta_\B = {405\Gamma_{\rm sph}\over 4\pi^2v_wg_*T}\int dz\, \mu_{B_L} 
	f_{\rm sph}\,e^{-45\, \Gamma_{\rm sph}|z|/(4 v_w)}
\label{baueq}
\end{equation}
where $f_{\rm sph}$ is a function that quickly goes to zero in the broken phase \cite{Cline:2011mm},
to model the spatial dependence of the sphaleron interaction rate, whose value in the symmetric phase is $\Gamma_{\rm sph} \cong 10^{-6}\, T$.  We take the wall velocity to be $v_w = 0.1$.  Since $\mu_{B_L}\sim v_w$ for small $v_w$, which cancels the factor of $1/v_w$ in (\ref{baueq}), 
our results are not very sensitive to the value of 
$v_w$ in the expected range $v_w\sim 0.01-0.1$. (For $v_w\ll 0.01$, the slowly moving wall does not provide enough departure from equilibrium to prevent washout by sphaleron interactions, as encoded by the exponential factor.)

\begin{figure}[t]
\vskip-0.3cm
\centerline{
\includegraphics[width=0.8\hsize]{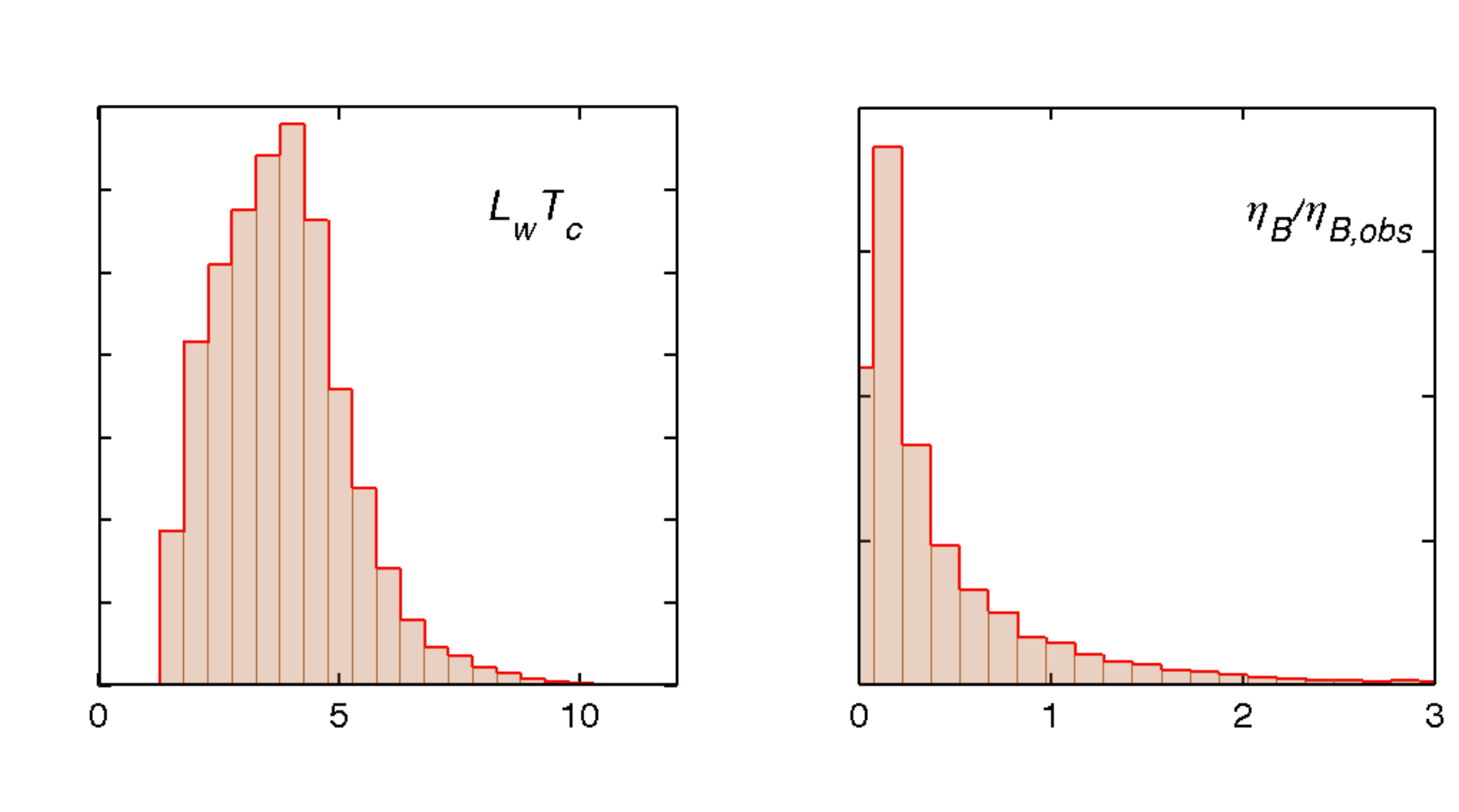}}
\caption{Distributions of wall thickness $L_w$, in units $T_c^{-1}$, and baryon asymmetry
$\eta_\B$, in units of the observed value.  The latter can also be read as the
distribution of $(\Lambda/{\rm TeV})^2$ for models with $\eta_\B=\eta_{\B,\rm obs}$.} 
\label{panel-bau}
\end{figure}

We solve the transport equations both by shooting and by relaxation
algorithms with consistent results. (We find that relaxation gives
somewhat more accurate solutions for the chemical potential profiles,
and we used this method for the results presented here.)
Of the randomly generated models passing all other constraints,
discussed in the previous sections, we find that 11\% have
$\eta_B$ exceeding the observed value, assuming the scale suppressing the
dimension-5 source of CP violation is fixed to be $\Lambda=1$ TeV. 
This demonstrates that it is relatively easy to get sufficient baryogenesis once the
phase transition is required to be strong enough ($v_c/T_c > 1$).  We
display the distributions of the wall thickness $L_w$, and $\eta_\B$
(in units of the observed value, $\eta_{\B,\rm  obs} = 8.7\times
10^{-11}$) in Fig.\ \ref{panel-bau}.  If we rescale $\Lambda$ so as to
give $\eta_\B$ the observed value, then Fig.\ \ref{panel-bau}(b) can
be reinterpreted as the distribution of values of $\Lambda^2$ in
TeV$^2$.

\begin{figure}[t]
\centerline{
\includegraphics[width=0.58\hsize]{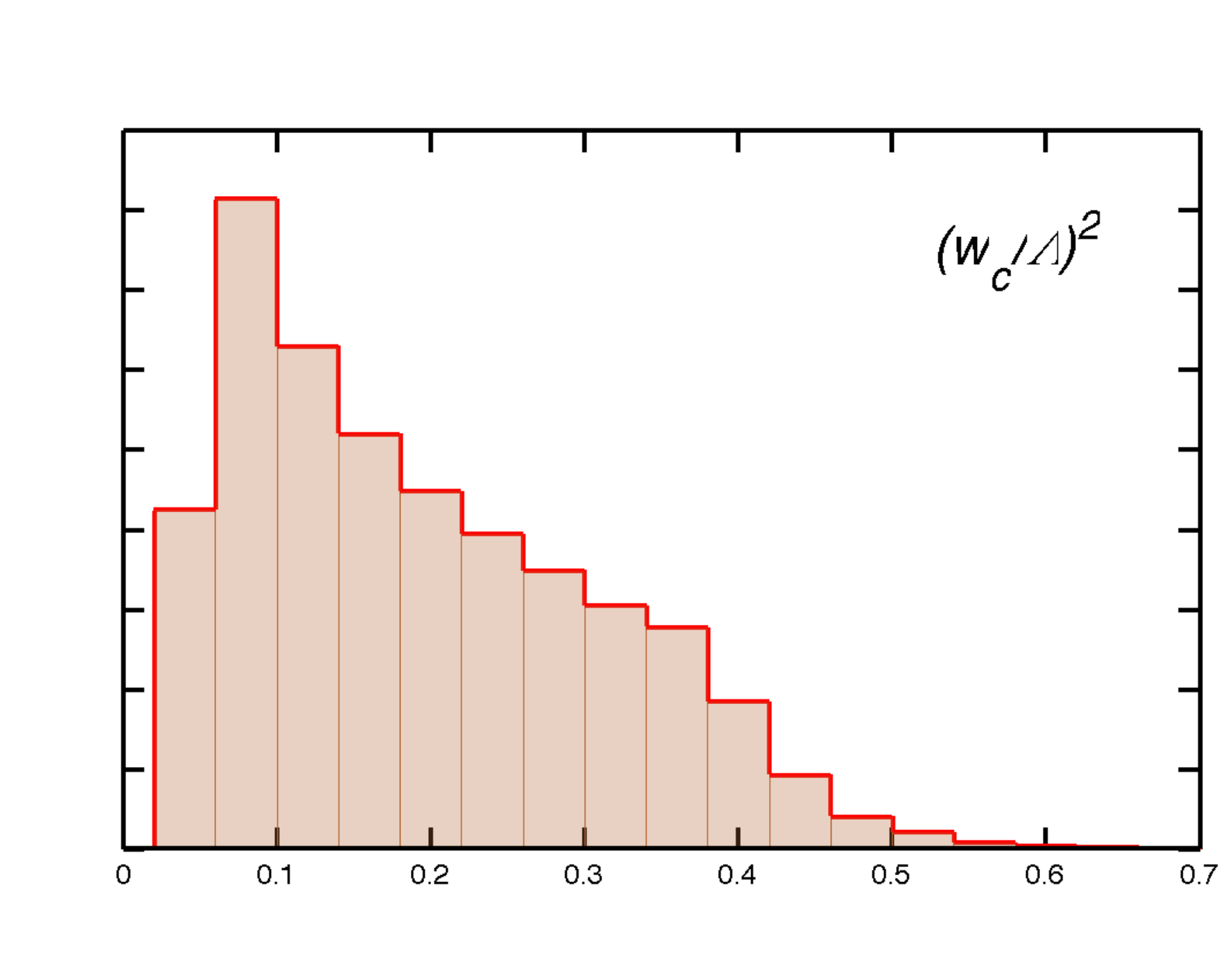}}
\caption{Frequency of $w^2_c/\Lambda^2$ from random scan of models, which should be small to justify the effective field theory treatment of the dimension-6 CP-violating operator.}
\label{hist-frac}
\end{figure}

We find that large values of the baryon asymmetry are correlated with
strong phase transitions, which is expected since the source of CP
violation depends partly upon the amount by which $|m_t|\sim v_c$
changes in the bubble wall. They also correlate with large values of
$w_c$ (the critical VEV of $S$) since this controls the imaginary part
of the top quark mass in the bubble wall. For consistency however, one
should demand that the dimension-6 contribution to $m_t(z)$ be small
compared to the renormalizable one, hence $(w_c/\Lambda)^2$ should be
small. After adjusting $\Lambda$ to make $\eta_\B = \eta_{\B,\rm
obs}$, we find that $w_c^2/\Lambda^2$ is distributed as shown in Fig.\
\ref{hist-frac}.  All accepted models have $w_c^2/\Lambda^2< 1$ and
29\% of them satisfy $w_c^2/\Lambda^2< 0.1$.

\begin{figure}[t]
\centerline{
\includegraphics[width=9cm]{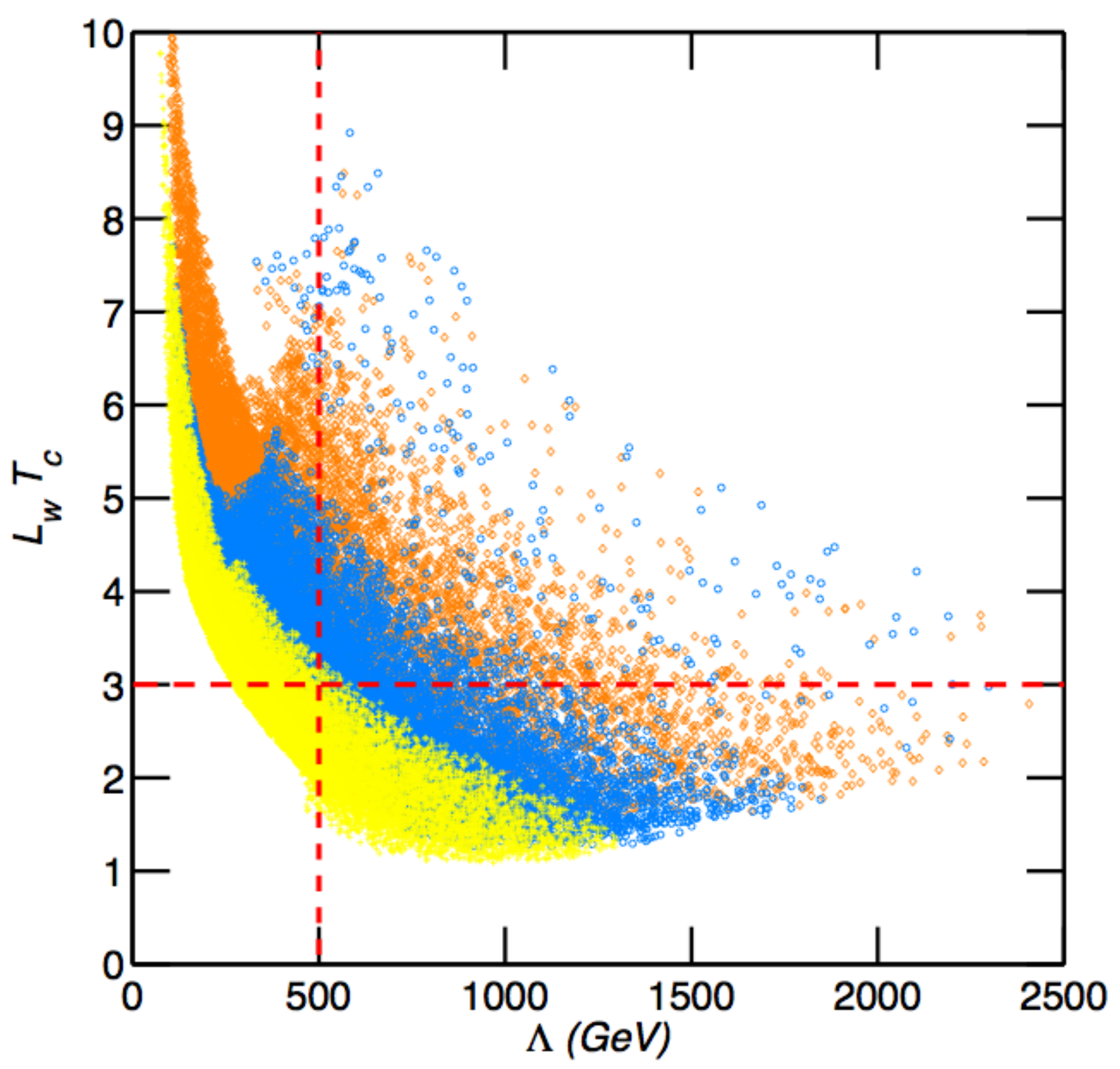}}
\caption{Scatter plot $\Lambda$ vs.\ $L_wT_c$ for the models 
consistent with constraints of sections  \ref{effpot} and 
\ref{higgsbound}.  Dashed lines indicate the consistency
constraints $\Lambda > 500$ GeV and $L_w> 3 T_c^{-1}$.
The colour-coding of points is the same as in figures~\ref{fpan} and~\ref{fpan2}.}
\label{lastscatter}
\end{figure}

\begin{figure}[t]
\centerline{
\includegraphics[width=\hsize]{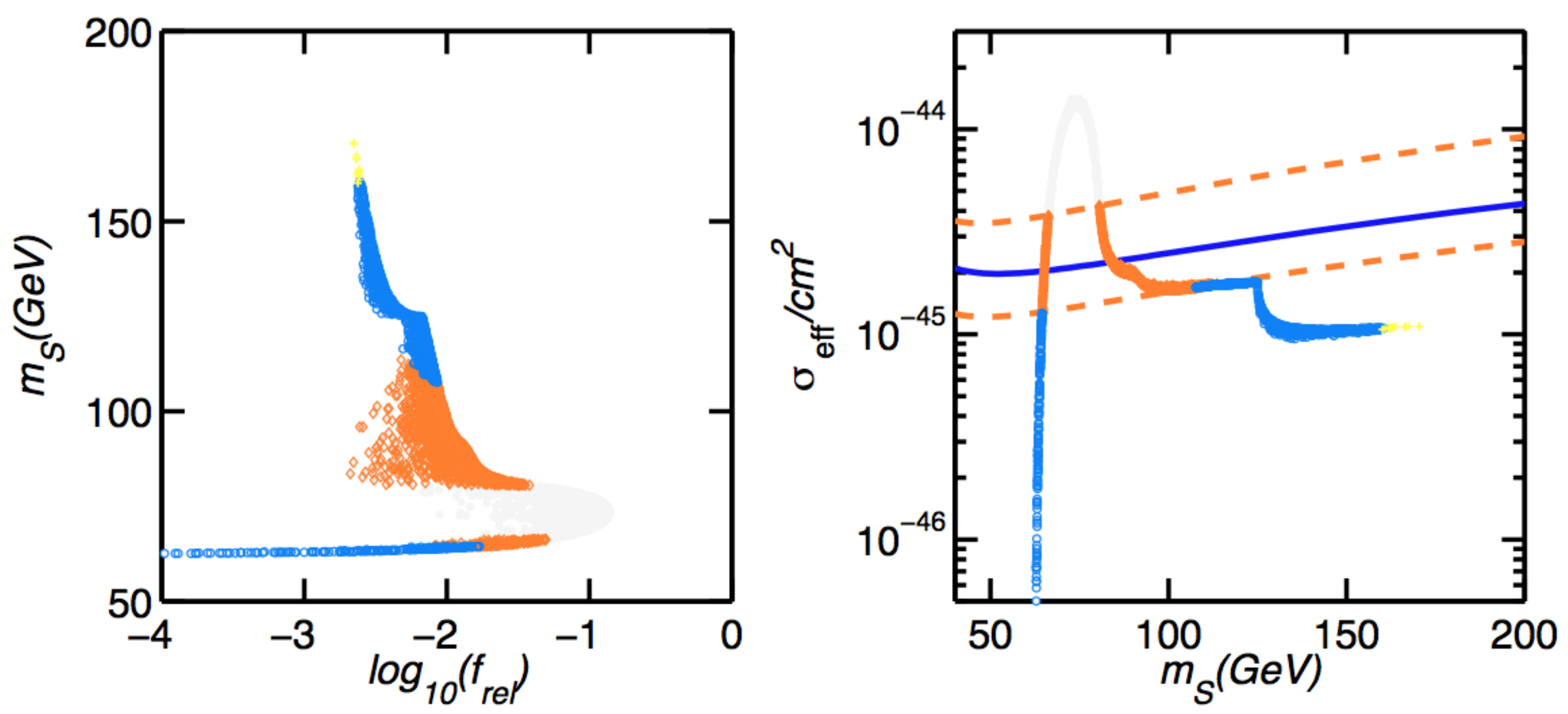}}
\caption{Scatter plots corresponding to figures~\ref{fpan}
and~\ref{fpan2}, showing all models that are also consistent with the further constraints $L_wT_c>3$ and $\Lambda>500$ GeV.}
\label{lastplot}
\end{figure}

From fig.\ \ref{panel-bau} we see that both $L_w$ and $\Lambda$ tend
to be somewhat small. These tendencies are actually correlated as is
evident from fig.~\ref{lastscatter}. This may be a limitation because
for very small wall thicknesses our baryogenesis formalism which
relies on the semiclassical
method~\cite{Joyce:1994zt,Cline:1997vk,Cline:2000nw,Kainulainen:2001cn,Kainulainen:2002th}
source becomes less reliable. To illustrate the tension, 
fig.~\ref{lastplot} displays the distribution of allowed models
subjected to additional constraints $L_wT_c>3$ and $\Lambda>500$ GeV.
Interestingly, these cuts tend to remove models 
with large $\lambda_m$, making the scenario even more
testable by direct DM searches. But even
after these stringent cuts there remains a
much greater fraction of parameter space
with viable examples than was found in the general two-Higgs doublet
model \cite{Cline:2011mm}, where a Markov-chain Monte Carlo approach
was  needed to find a significant number of working models.

\bigskip
\section{Conclusions}
\label{conc}

In this paper we have established a new possible connection between
electroweak baryogenesis and dark matter, via a singlet Higgs particle
$S$.  Even though $S$ cannot constitute the majority of the dark
matter in this scenario (at most $\sim 3$\%), it may still be on the
verge of discovery by XENON-like direct detection searches, due to its
large Higgs-mediated cross section on nucleons. Our predicted range of
masses $m_\S = 80-160$ GeV (for the case where $\lambda_m$ is
restricted to be  $< 1$) is near that where XENON100 is most
sensitive.

Apart from providing a new dark matter candidate, this theory is
interesting in that it is easy to find parameters giving a sufficient
baryon asymmetry, thanks in part to the mechanism of getting a strong
electroweak phase transition from having a large potential barrier at
tree level, that allows an analytic treatment of the phase transition
properties.  The model is not UV-complete because we have invoked a
nonrenormalizable operator $\bar Q_\L H S^2 t_\R$ to get CP
violation.  However such an operator could be obtained from a
renormalizable theory by integrating out a heavy Higgs doublet $H'$
that does not get a VEV, has a Yukawa coupling to the top quark and a
(possibly complex) $H^\dagger H' S^2$ quartic coupling.  We found that
the new physics scale $\Lambda$ could be sufficiently large to justify the low-energy
effective field theory treatment.

It would be interesting to repeat this exercise in the context of
two-Higgs doublet models, where the second doublet is not very heavy.
In the purely numerical search done in ref.\ \cite{Cline:2011mm},
there is not much insight into what is special about the models that
work (other than that large values of $S$ are needed in the bubble
wall to get enough baryon asymmetry, in common with the present
situation).  It is possible that they are variants of the mechanism
used here, involving the large tree-level potential barrier. Moreover,
the studies of the Inert Doublet Model done in ref.\
\cite{Chowdhury:2011ga,Borah:2012pu,Gil:2012ya} could likely be generalized to
find a similar DM mass range as we found here, if the possibility of
subdominant DM and all the relevant annihilation channels were
included.

It is intriguing that the DM mass range we found includes 130 GeV, since there are exciting indications of DM annihilation into two photons of approximately this energy \cite{Weniger:2012tx}-\cite{Cohen:2012me}.  Recently it was proposed that subdominant DM could account for this observation \cite{Cline:2012bz}. In that case it was necessary to have both a larger fraction of DM in the 130 GeV component, and for the dominant annihilation channel to be into two photons, neither of which are the case in the present scenario.  However these conclusions may not necessarily hold for other models, and so it could be worthwhile to explore whether the phenomena we focus on here could be compatible with a simultaneous explanation of the 130 GeV line.

\bigskip
{\bf Acknowledgments}
The authors thank the CERN Theory Division for its hospitality during part of this work.  JC is supported by the Natural Sciences and Engineering Research Council (NSERC, Canada).

\bibliographystyle{apsrev}

\end{document}